\documentclass{INTERSPEECH2023}
\usepackage{multirow}  
\usepackage{cite}

\title{EDTC: ENHANCE DEPTH OF TEXT COMPREHENSION IN AUTOMATED AUDIO CAPTIONING}
\name{Liwen Tan$^1$$^,$$^2$, 
    Yin Cao$^3$, 
    Yi Zhou$^1$$^,$$^2$$^,$$^*$\thanks{*Corresponding author}}
\address{
    School of Communication and Information Engineering, Chongqing University of Posts and Telecommunications, Chongqing (CQUPT), China$^1$\\
    Intelligent Speech and Audio Research Lab (ISARL), CQUPT, Chongqing, China$^2$ \\
    Department of Intelligent Science, Xi’an Jiaotong-Liverpool University, China$^3$}
\begin{document}

\maketitle

\begin{abstract}

Modality discrepancies have perpetually posed significant challenges within the realm of Automated Audio Captioning (AAC) and across all multi-modal domains. Facilitating models in comprehending text information plays a pivotal role in establishing a seamless connection between the two modalities of text and audio. While recent research has focused on closing the gap between these two modalities through contrastive learning, it is challenging to bridge the difference between both modalities using only simple contrastive loss. This paper introduces Enhance Depth of Text Comprehension (EDTC), which enhances the model's understanding of text information from three different perspectives. First, we propose a novel fusion module, FUSER, which aims to extract shared semantic information from different audio features through feature fusion. We then introduced TRANSLATOR, a novel alignment module designed to align audio features and text features along the tensor level. Finally, the weights are updated by adding momentum to the twin structure so that the model can learn information about both modalities at the same time. The resulting method achieves state-of-the-art performance on AudioCaps datasets and demonstrates results comparable to the state-of-the-art on Clotho datasets.

\end{abstract}

\section{Introduction}

\begin{figure}
    \centering
    \includegraphics[width=1\linewidth]{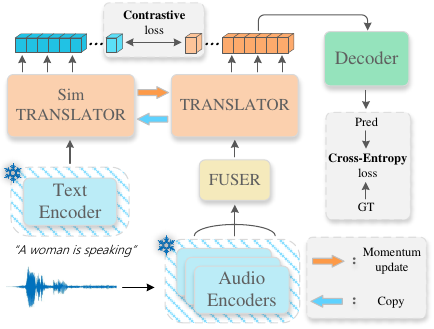}
    \caption{Overview of the methodology presented in this paper. Audio encoder uses PANNs\cite{kong2020panns}, HTSAT\cite{chen2022hts}, ConvNeXt\cite{liu2022convnxet} for extracting audio features. INSTRUCTOR-base\cite{su2022Instructior} is used as Text Encoder for extracting text features and BART is used as Decoder for generating captions. All encoder weights are frozen.}
    \label{fig:1}
\end{figure}

Automated Audio Captioning (AAC) is a multimodal generation task that integrates audio and text to generate descriptive captions for audio\cite{mei2022automated}. The applicability of AAC research is diverse, encompassing areas such as aiding the hearing-impaired through audio-to-text conversion and building intelligent content-oriented human-computer interactions\cite{xu2022comprehensive}. AAC research has received a dramatic increase in attention and made substantial progress in recent years, thanks to the DCASE challenge and the release of datasets such as Clotho\cite{drossos2020clotho} and AudioCaps\cite{kim2019audiocaps}.

The current mainstream AAC methods employ encoder-decoder system architectures. Specifically, an audio encoder such as PANNs\cite{kong2020panns}, HTSAT\cite{chen2022hts}, etc., is utilized to extract the acoustic features of log-mel energies or Mel-Frequency Cepstral Coefficients (MFCCs), and then the text decoder generates the textual output. Text decoders generally use sequence-to-sequence design. In recent years, with the development of Transformers architecture, more and more excellent decoders such as GPT\-2\cite{radford2019language}, BART\cite{lewis2019bart}, etc., have gradually become the first choice for text decoders.

In previous studies, many researchers have explored AAC research through various perspectives. Some researchers have improved system performance by introducing additional feature embeddings (such as sound event detection\cite{xie2023sed} or keyword information\cite{eren2023topic}). Other researchers directly optimize the metrics to enhance the quality of generated subtitles by using reinforcement learning algorithms\cite{mei2021rl}. In order to solve the problem of scarce training data, Mei et al. proposed WavCaps\cite{mei2023wavcaps} dataset, and Wu et al. proposed LAION-Audio-630K dataset\cite{wu2023clap}, which greatly solved this problem. Despite these approaches, a fundamental challenge remains unresolved — the huge disparity between the audio and text modalities.

With the popularity of Contrastive Language-Image Pretraining (CLIP)\cite{radford2021clip} in image captioning, contrastive learning has gradually gained attention from researchers. CLIP extracts the features of two modalities through two independent encoders and then closes the gap between the two modalities by using a contrastive loss. Elizalde et al. proposed Contrastive Language-Audio Pretraining (CLAP) to apply CLIP to audio-text domain by replacing the image encoder in CLIP with an audio encoder\cite{elizalde2023clap}. Wu et al. applied CLAP to AAC research by large scale training\cite{wu2023clap}. Chang et al. used a similar architecture but used InfoNCE loss to close the gap between the two modalities\cite{wu2023SOTA}. Despite the fact that all the referenced studies yielded outstanding outcomes, they uniformly employed a contrastive loss function to facilitate the learning of text features by the audio encoder, derived from the text encoder, without fully capitalizing on the information extracted by the text encoder. This raises a critical question: Is it feasible for a mere loss function to adequately bridge the substantial gap between the audio and text modalities?

Previous research often encountered a significant challenge: directly feeding audio features into the text decoder in an attempt to directly interpret audio features through the text decoder, a task that proves to be exceedingly difficult. Consider an analogy where the audio information represents "meow" and the text information signifies "bark"; even with the most intelligent cats and dogs, is mutual comprehension truly achievable? To bridge the gap between these two distinct modalities, it is essential to employ auxiliary techniques that facilitate the convergence of the audio and text modalities within the model.

In this paper, we propose to use three different approaches in order to address the disparity between the audio and text modalities. The details of each method and the reason why it was chosen are described in Section 3. Our model, shown in Figure 1, is divided into audio and text branches. The audio branch first extracts audio features by multiple different audio encoders with frozen weights, and then a feature fusion module FUSER fuses the extracted audio features to extract the shared semantic information. Then, the fused features are fed into an alignment module TRANSLATOR for feature alignment in the tensor level. Finally, the output features are fed into a text decoder to get the generated captions. The text branch consists of a frozen text encoder and a twinned TRANSLATOR module. First, the text encoder extracts text features from the original captions, and then the extracted text features are passed through a Sim TRANSLATOR module with the same structure as the audio branch, the TRANSLATOR modules of the two branches share the weights and use momentum to update the weights in such a way that the TRANSLATOR module of the audio branch is also able to learn the text features directly. Finally, a contrastive loss assist model bridge the difference between both audio and text modalities. In general, the contributions of this paper are as follows:

\begin{itemize}

\item We propose a novel feature fusion module, FUSER, which can extract shared semantic information from the output features of multiple different audio encoders and act as a plug-and-play feature extractor to extract higher dimensional semantic information when input to multiple identical encoders.

\item We propose a new plug-and-play alignment module, TRANSLATOR, to further bridge the difference between the two modalities by aligning audio features with text features in the tensor level.

\item We propose to use a set of twin TRANSLATOR structures, combined with the momentum update weight and the addition of auxiliary contrastive loss to enable the model to directly learn the information of the two modalities simultaneously.

 \end{itemize}

\section{Related work}

This section will provide a brief overview of the research areas related to this paper, including audio captioning, feature fusion, contrastive learning.

\subsection{Audio captioning}

The aim of audio captioning is to generate descriptive captions for audio. Many researchers currently employ transfer learning methods to allow models to learn more samples in order to enhance model performance. Reinforcement learning is also used in this field for direct optimization of metrics, However, it will cause the generated captions to be not fluent. Wu et al. applied CLAP to AAC research by large scale training\cite{wu2023clap}, and Chen et al. proposed CLIP-AAC by a similar approach\cite{chen2022clip-aac}, where by extracting the features of their respective modalities through two different modality encoders, and then through clip loss (a kind of contrastive loss) to bridge the difference between the two modalities. Chang and Wu et al. adopted a similar architecture, but used BAETs + Conformer as the audio encoder and employed InfoNCE loss instead of clip loss\cite{wu2023SOTA}. Deshmukh et al. attempted training the model using only text information based on this architecture\cite{deshmukh2023without_audio}, which proves that this architecture can indeed bridge the difference between both modalities effectively. The method proposed in this paper is to learn information about both modalities simultaneously through a set of twinned structures, instead of bridging the difference between both modalities with only a single contrastive loss. Furthermore, we propose to further aid the model in understanding the disparity between modalities through feature fusion and tensor level alignment.

\subsection{Feature fusion}

The aim of feature fusion is to integrate the information of different features, which can be usually divided into the same modality feature fusion and multi-modal feature fusion. Based on the model architecture, it can be classified into the use of a unified architecture model and the use of a non-unified architecture model. Using a unified architecture model for fusion\cite{jung2020unified1, xu2020unified2, xu2020unified3}, that is, different features are fused with the same architecture model. This kind of method can effectively alleviate the problem of limited data amount by using different kinds of data (such as audio, text, image, etc.) for training. Which can effectively extract the common information between features, but the unique information of each feature is often ignored. The method of using no unified architecture model\cite{kuo2023non-unified1, huang2019non-unified2, jiang2018non-unified3}, that is, first using a separate model for each feature to extract features, and then performing some series of operations such as averaging or concatenation or cross-attention on the extracted features to obtain the final features. Which can effectively extract the unique information of various features, but exhibits limited capability in capturing common information. Our proposed method integrates both approaches. Firstly, audio features are extracted through different feature encoders, because different encoders can extract features from different perspectives for the same audio sample (encoders based on CNN architecture pay more attention to local features, encoders based on Transformer architecture are better at extracting global information). Subsequently, these features are fed into a fusion module, where common and unique information is separately extracted during the fusion process. The amalgamation of these pieces of information generates higher-dimensional semantic information, thus narrowing the gap between the two modalities of audio and text.

\subsection{Contrastive learning}

Contrastive learning has long captivated researchers owing to its advantages of self-supervised learning, and its goal is to extract useful information by learning the differences and similarities between samples. In previous studies, researchers obtained different views of the same image by applying different data enhancement methods to the same image\cite{he2020MoCo, chen2020simcle}. Then compare the different views to get useful information. In recent studies, a series of studies on contrastive learning, mainly represented by CLIP\cite{radford2021clip}, have gradually become popular in multimodal tasks. There are also many researchers in the field of AAC have done similar work\cite{wu2023clap, chen2022clip-aac, deshmukh2023without_audio}, they basically use the same architecture. The model is divided into two branches, each branch has an encoder for extracting features, and then the contrastive loss is used to bridge the difference between the features of the two modalities. While this approach is straightforward, it presents some challenges. As mentioned earlier, the large disparity between the two modalities is difficult to accommodate with a single loss function. In addition, this will inevitably introduce an additional text encoder to extract text features. If only the text features and audio features are used to calculate the loss, this will not make full use of text features and cause a waste of resources. In this study, we add a set of twin modules to the audio branch and the text branch, and update the weight by momentum, so that the modules of the two branches can share the weight and learn the features of their respective branches, so as to effectively use the text features. This also allows the audio encoder to learn the text features directly, instead of just bridging the difference between two features via the loss function.

\section{Method}

In this section, we will provide a comprehensive description of the methodology proposed in this paper, including a detailed rationale for its selection.

\subsection{Encoders and decoder}

There are many excellent encoders that have been applied to AAC in previous studies\cite{kong2020panns, chen2022hts, liu2022convnxet}. These encoders are based on different architectures, which gives them different perspectives when extracting features. For example, PANNs\cite{kong2020panns} is based on CNN architecture, and its extracted features are more local detailed features, while HTSAT\cite{chen2022hts} based on Transformer architecture will focus more on global information due to the application of attention mechanism. The proposed method aims to extract common semantic information by integrating features from different views.

\subsubsection{Audio encoders}

We choose three widely used Pre-trained audio encoders, PANNs\cite{kong2020panns}, HTSAT\cite{chen2022hts} and ConvNeXt\cite{liu2022convnxet}, to extract audio features under various views. PANNs\cite{kong2020panns} is based on CNN architecture and consists of 14 convolutional layers. HTSAT\cite{chen2022hts} is an audio encoder based on the Swin-Transformer architecture. ConvNeXt\cite{liu2022convnxet} is also based on CNN architecture, but different from PANNs, it uses larger convolution kernels and reference part of Transformer structure settings. All three audio encoders are frozen with weights and do not participate in weight updates.The audio feature tensors extracted by each audio encoder are resized to the same shape of 32 × 768.

\subsubsection{Text encoder}

We adopt INSTRUCTOR-base\cite{su2022Instructior} as the text encoder to extract text features from captions. Which is a pre-trained T5 model based on instruction fine-tuning. During training, the weights of INSTRUCTOR are frozen and use "Represent the audio caption:" as the instruction to generate word-level text embeddings. The tensor shape out of the final extracted text embedding is adjusted to the same shape as the audio encoders.

\subsubsection{Text decoder}

 We adopt BART\cite{lewis2019bart} as the text decoder, which is a Transformer decoder with bidirectional encoder and autoregressive decoder (BERT + GPT). Use the same parameter settings as BART-base in Hugging Face, which have 6 layers of encoder and 6 layers of decoder, and have a 5K vocabulary size. Unlike the encoder part, BART takes part in the training and updates the weights.

Cross-entropy loss is the main loss function of the model, which is obtained by calculating the output distribution of BART and the true captions. The cross-entropy loss can be expressed as follows:
\[L_{ce} = -\frac{1}{N} \sum_{n=1}^{N} \log{p\left ( y_{n} \mid y_{1:n-1};x  \right ) } \tag{1}\]
Where \(x\) is the audio sample, \(y\) is the audio caption, \(y_{n}\) is the \(n^{th}\) word in the caption, and \(N\) is the batch size.
 
\subsection{FUSER}

\begin{figure}
    \centering
    \includegraphics[width=1\linewidth]{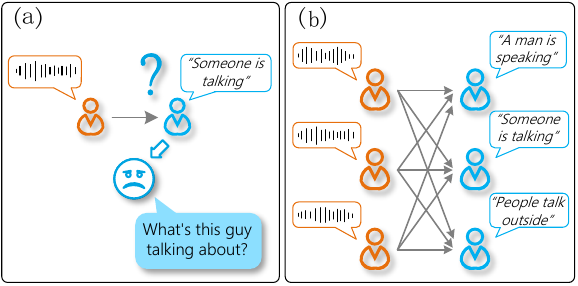}
    \caption{The statement about the reason for selecting FUSER, where the orange dialog box represents the audio sample, and the blue dialog box represents the text corresponding to each piece of audio.}
    \label{fig:enter-label}
\end{figure}

This section describes the FUSER module structure in detail and why it was chosen.

\begin{figure*}[t]
    \centering
    \includegraphics[width=1\linewidth]{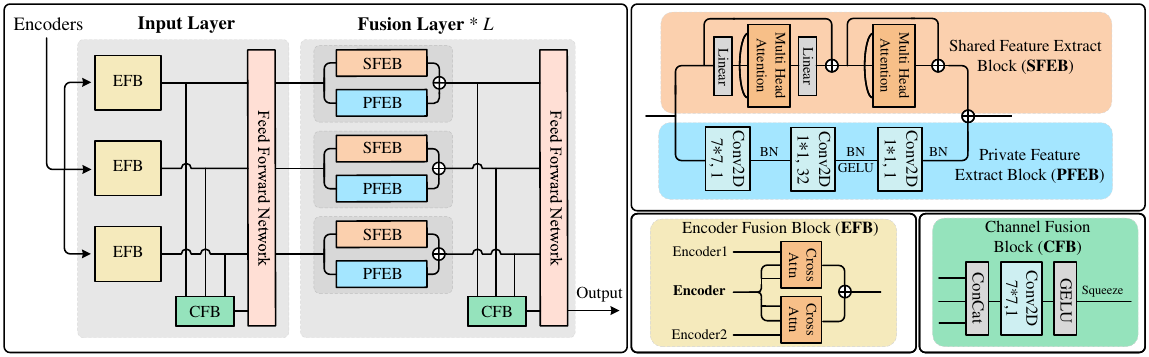}
    \caption{Detail diagram of the FUSER module. The module consists of an Input layer for the pre-fusion and a Fusion layer for the L layers. The input of the module is the audio features extracted for each encoder, and the output is the final fusion features.}
    \label{fig:enter-label}
\end{figure*}

\subsubsection{Why choose FUSER}

To put it another way, the goal of AAC task is to make the text decoder understand the audio features extracted by the audio encoder. If the interaction between an audio encoder and a text decoder is compared to a conversation between two people, as shown in Figure 2, where one person (the audio encoder) speaks English and the other (the text decoder) speaks French, it is obvious that no matter how the English speaker condense his expression, it will be difficult for the other person to understand because these are two different languages (modalities) anyway. To solve this problem, now let three different people (three different audio encoders) express the same information, and let three different people (the fusion structure) each understand all the different expressions. Although a single person still cannot understand each other, from different expressions and different understandings, some common points can always be found. These common points are the so-called semantic information, which does not belong to any single modality. In this way, a bridge can be built between the two modalities to help them understand each other. FUSER, proposed in this paper, aims to extract semantic information from audio features by this method.

\subsubsection{FUSER specific details}

The specific structure diagram of FUSER is shown in Figure 3, which consists of an Input Layer and multiple repeated Fusion layers. The inputs of the FUSER are audio features extracted from three different audio encoders. First, these features are fed into an Input Layer, which consists of three Encoder Fusion blocks (EFB), a Channel Fusion Block (CFB) and a Feed Forward Network (FFN). Each EFB will take one audio feature as the baseline, compute the cross-attention and sum it with the other two features. The outputs of each EFB are then concatenated together and use a convolutional layer with a kernel size of 7 for dimension reduction. Finally, the output of each EFB and the output of the CFB are fed into a shared weight FFN to obtain the output. Through the Input Layer, the features are fused in advance before deep fusion to improve the fusion quality.

The structure of the Fusion Layer is similar to that of the Input Layer, but the EFB module is replaced by Shared Feature Extract Block (SFEB) and Private Feature Extract Block (PFEB). PFEB is used to extract the unique information of each feature. This kind of information tends to be local features. We use a 3-layer convolution architecture to extract these local features, and add Batch Normalization between each layer to reduce overfitting. SFEB is used to extract the common information of each feature, which is usually global information. We extract the global information by calculating the self-attention of the channel dimension and the feature vector dimension, and the weight is shared between each SFEB. In order to reduce memory consumption, the linear layer is used to reduce the dimension from 768 to 256 before calculating the self-attention of the feature vector dimension, and then the dimension is re-increased to 768 by the linear layer after the calculation. The features will be deeply fused through the L-layer Fusion Layer to finally obtain the fused features. We experimentally find that L = 3 performs best.

It is worth noting that the input of FUSER can be replaced with the features extracted by any of the three encoders. And when the model structure has only one encoder, three same features can also be used as the input of FUSER. At this time, the situation is similar to one person express and three people understand, and they also have the ability to extract semantic information from the same feature.

\subsection{TRANSLATOR}

This section describes the TRANSLATOR module structure in detail and why it was chosen.

\begin{figure}
    \centering
    \includegraphics[width=1\linewidth]{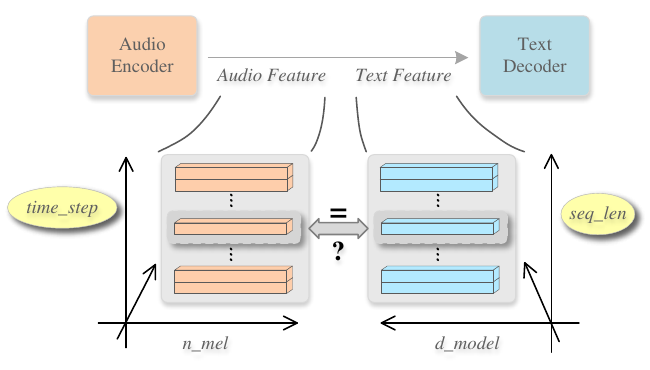}
    \caption{The statement about the reason for selecting TRANSLATOR, where time\_step is the time step, n\_mel is the number of mel filters, seq\_len is the text length, d\_model denotes the encoding dimension of each word.}
    \label{fig:enter-label}
\end{figure}

\subsubsection{Why choose TRANSLATOR}

Generally, in the construction of the model, different modules can be spliced together by just adjusting the input and output tensors to the same shape, but this completely ignores the specific significance of each dimension of the tensor. As shown in Figure 4, in AAC task, audio features are basically obtained by extracting the corresponding log mel spectrum, and the shape of audio feature tensor is (bs, time\_step, n\_mel), where bs stands for batch size, time\_step stands for time step, and n\_mel stands for the number of mel filters. The text features is represented by a tensor of shape (bs, seq\_len, d\_model), where bs is the batch size, seq\_len is the text length, and d\_model is the encoding dimension of each word. By comparison, we can see that only the first dimension of the two tensors represents the same meaning, and the last dimension should also represent a similar meaning through complex changes of the model, but the second dimension is difficult to establish a corresponding relationship. Let's imagine that each time step of the audio features can really correspond to every word in the text?

It can be imagined that the correspondence between each time step and each word is not a simple one-to-one relationship, but a many-to-many relationship. For text information, each word contains not only its own information, but also more context information, while for audio samples, each time step more focuses on the current time node information. By making the features at each time step contain more context information, we align the two features in the tensor level to reduce the gap between the two modalities, which is the specific role of the TRANSLATOR module.

\begin{figure}
    \centering
    \includegraphics[width=1\linewidth]{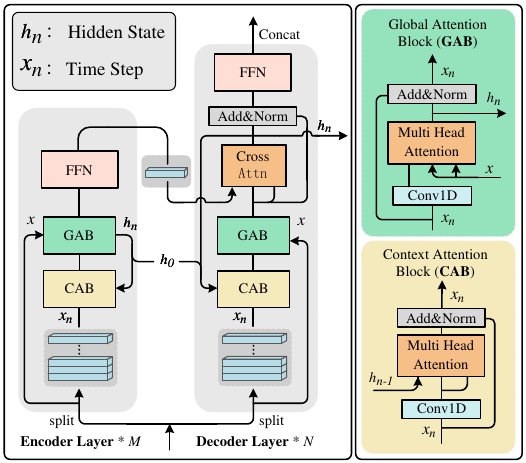}
    \caption{Detail diagram of the TRANSLATOR module, which is composed of an M-layer Encoder Layer and an N-layer Decoder Layer. The module input is split into individual time steps, and then the output of each time step is concatenated into a complete output after passing through the module.}
    \label{fig:enter-label}
\end{figure}

\subsubsection{TRANSLATOR specific details}

As shown in Figure 5, TRANSLATOR is composed of M-layer encoder and N-layer decoder, whose input is a tensor of shape (bs, time\_step, d\_model), and splits it in the time\_step dimension. Each split tensor has the shape (bs, d\_model) and is used as input for each time step.

The encoder consists of a Context Attention Block (CAB), a Global Attention Block (GAB) and a standard Feed Forward Network (FFN). Inside CAB, the input is first extracted by a Conv1D with a kernel size of 45 to extract the current time node features. Through this step, the input which is originally a whole can be subdivided into a single time step instead of simply splitting the tensor. Using a large convolution kernel can reduce the influence of convolution on the original information at each time step by expanding the receptive field. After the convolutional layer, the cross-attention information of the current time step and the hidden state of the previous time step are calculated. Finally the output of the CAB module is obtained by layer normalization through a residual structure. The feature of the current time step through the CAB module will carry the information of all previous time steps. After that, each time step feature passes through a GAB module, which has the same internal structure as CAB, but computes cross-attention with the original input that was not split. The value obtained through the calculation is passed to the next time step as the hidden state of the current time step. Through the GAB module, the feature at the current time step further have global context information. Finally, the feature are linearly transformed by an FFN to obtain the output of the encoder at the current time step.

The structure of decoder is similar to encoder, but a part of the structure between GAB module and FFN is added to calculate the cross-attention of encoder feature and decoder feature at the same time step, so that it can have more abundant global information. The calculated value is used as the hidden state of decoder at the current time step. In addition, the last hidden state of the last layer of the encoder is used as the initial hidden state at time step 0 of each layer of the decoder. Finally, the output of each time step of decoder is spliced together as the final output of TRANSLATOR. We experimentally find that M = 3, N = 2, and kernel\_size = 45 perform best.

The output of the TRANSLATOR \(y_{tran}\) can be expressed as follows: 
\[{\Huge x_{0},\ x_{1},  \cdots\ x_{T} = Split\left ( X \right )} \tag{2} \]
\[y_{tran} = Concat_{n=0}  ^{T}\  f\left ( x_{n} \ \odot \ h_{n-1}\right ) \tag{3}\]
Where \(X\) represents the TRANSLATOR's input, \(x_{n}\) represents the input at  \(n^{th}\) time step, \(h_{n-1}\) represents the hidden state at \(\left ( n-1 \right ) ^{th}\) time step, \(T\) represents the total number of time steps,  \(f\left ( \cdot  \right ) \) represents the TRANSLATOR's calculation process, \(split\left ( \cdot  \right ) \) represents the tensor split along the time dimension, \(Concat\left ( \cdot  \right ) \) represents the tensor resplicing, \(\odot \) stands for computing cross-attention. In summary, TRANSLATOR aligns the text information in the tensor level by performing an identity transformation on the audio features. It can be added as a plug-and-play alignment module to almost all models in the AAC field.

\subsection{Twin structure and momentum update}

This section describes the twin structure and momentum update in detail and why it was chosen.

\subsubsection{Why choose this structures}

As a multi-modal task, how to make full use of text information has always been one of the key points to improve the performance of AAC system. A most direct approach is to use audio and text as the input of the model at the same time. In this way, the information of audio and text  can be directly used at the same time. while since only audio information can be used in the inference process, this method will make the model miss part of the input during inference. Previous studies almost bridge the difference between both modalities by using contrastive loss\cite{elizalde2023clap, wu2023clap, wu2023SOTA, chen2022clip-aac, deshmukh2023without_audio}. Although this method is simple and easy to use, it is an indirect use of text information. Isn't there a way to use text information directly? In order to use text information directly, a core problem must be overcome, which is how to make the model learn how to extract text features while retaining the ability to extract audio features.

\begin{table*}[]
\large
\caption{Performance in Clotho and AudioCaps\cite{kim2019audiocaps} evaluation split, where Extra training data denotes the use of additional data during the training phase. AC denotes AudioCaps and WC denotes WavCaps\cite{mei2023wavcaps}. Bold indicates the best result for each group.}\label{tab:tablenotes}
\renewcommand\arraystretch{1.3}
\setlength{\tabcolsep}{1mm}{\resizebox{\textwidth}{!}{
\begin{tabular}{c|c|c|ccc|ccccc}
\hline
\textbf{Dataset}           & \textbf{Methods}   & \textbf{Extra training data} & \textbf{BLEU\(\mathbf{_{1}}\)}   & \textbf{BLEU\(\mathbf{_{4}}\)}   & \textbf{ROUGE\(\mathbf{_{l}}\)} & \textbf{METEOR} & \textbf{CIDER} & \textbf{SPICE} & \textbf{SPIDEr} & \textbf{SPIDEr\_FL} \\ \hline
\multirow{7}{*}{Clotho}    & Zhang et al.\cite{zhang2023zhang}       &  \textbf{-}         & 0.566            & 0.161            & 0.375            & 0.176          & 0.409          & 0.121          & 0.265          & \textbf{-} \\
                           & Ye et al.\cite{ye2021ye}          &   \textbf{-}        & 0.577            & 0.174            & 0.377            & 0.174          & 0.419          & 0.119          & 0.269          & \textbf{-} \\
                           & Komatsu et al.\cite{komatsu2023Komatsu}     & ESC-50              & \textbf{0.601}   & 0.166            & 0.388            & \textbf{0.194} & 0.454          & 0.127          & 0.291          & 0.289            \\
                           & Wu et al.\cite{wu2023SOTA} (SOTA)   & AC(ChatGPT assist)  & \textbf{-}       & \textbf{-}       & \textbf{-}       & 0.187          & 0.474          & 0.134          & 0.304          & 0.303            \\ \cline{2-11} 
                           & Ours               &  \textbf{-}         & 0.597            & \textbf{0.177}   & \textbf{0.391}   & 0.188          & \textbf{0.483} & \textbf{0.135} & \textbf{0.309} & \textbf{0.304}   \\ \cline{2-11} 
                           & Mei et al.\cite{mei2023wavcaps}         & WC                  & 0.601            & 0.180            & 0.400            & 0.185          & 0.488          & 0.133          & 0.310          & \textbf{-} \\
                           & Wu et al.\cite{wu2023SOTA} + Rerank & AC(ChatGPT assist)  & \textbf{-}       & \textbf{-}       & \textbf{-}        & 0.193          & 0.506          & 0.146          & 0.326          & 0.326            \\ \hline
\multirow{6}{*}{AudioCaps} & Liu et al.\cite{liu2022liu}         &  \textbf{-}         & 0.698            & 0.281            & 0.494            & 0.237          & 0.711          & 0.172          & 0.442          & \textbf{-} \\
                           & M. Kim et al.\cite{kim2023MKim}      & \textbf{-}          & 0.713            & \textbf{0.309}   & 0.503            & 0.240          & 0.733          & 0.177          & 0.455          & \textbf{-} \\
                           & Deshmukh et al.\cite{deshmukh2023Deshmukh}    &   \textbf{-}        & 0.691            & 0.253            & 0.482            & 0.232          & 0.752          & \textbf{0.182} & 0.467          & \textbf{-} \\
                           & E. Kim et al.\cite{kim2023EKim}      &   \textbf{-}        & 0.700            & 0.289            & 0.502            & 0.242          & 0.769          & 0.181          & 0.475          & \textbf{-} \\
                           & Mei et al.\cite{mei2023wavcaps} (SOTA)  & WC                  & \textbf{0.707}   & 0.283            & \textbf{0.507}   & 0.249          & 0.787          & \textbf{0.182} & 0.485          & \textbf{-} \\ \cline{2-11} 
                           & Ours               &  \textbf{-}         & 0.702            & 0.280            & \textbf{0.507}   & \textbf{0.254} & \textbf{0.806} & 0.180          & \textbf{0.493} & \textbf{0.484}   \\ \hline
\end{tabular}}}
\end{table*}

\subsubsection{specific details}

In the proposed method, the model is divided into audio branch and text branch. The detail of the model is shown in Figure 1. Firstly, each of the two branches extracts the features of the corresponding modality. In the audio branch, the combination of Encoders + FUSER is used to extract audio features. In the text branch, a frozen language model INSTRUCTOR-base is used to extract word-level features. Then, the features of each branch are input into a twin TRANSLATOR structure, which is composed of two identical TRANSLATOR modules with shared weights, and features of the two modalities are learned through this identical network structure. Although this method allows the TRANSLATOR to learn the features of two modalities at the same time, it still cannot solve the problem of missing text data in the inference stage. Therefore, we add two methods to solve this problem: (1) Updating the weight with momentum. (2) Adding contrastive loss.

Assuming that the weight of the TRANSLATOR in the audio branch is \(W\), and the weight of the TRANSLATOR in the text branch is \(W_{sim}\), the momentum update weight can be expressed as follows:
\[W = \beta \cdot W \ + \ \left ( 1-\beta  \right )\cdot  W_{sim}\tag{4}\]
Where \(\beta\) is the momentum parameter, which is used to control the proportion of updates at each step. Each time the model updates the weights, the twin TRANSLATOR structure is updated by using the momentum to update the weights. Specifically, after each weight update, the weight of the TRANSLATOR module in the audio branch is composed of two branch weights combined in a certain proportion. By increasing the weight proportion of the audio branch (\(\beta > 0.9\)), the ability of the audio branch TRANSLATOR to extract audio features is guaranteed while slowly learn text features. After the weights are updated, the weights of the TRANSLATOR in the audio branch are directly copied to the TRANSLATOR in the text branch. This is to ensure that the weights of the two branches are consistent, so as to prevent the weights of the two branches from being completely uncorrelated due to long training, which will lead to twin structure loses the ability to learn two modalities information at the same time. We experimentally find that \(\beta = 0.95\) is best.

There are two purposes for applying contrastive loss: (1) By distinguishing positive and negative sample pairs in each batch, the audio branch module can indirectly learn the text features extracted by the text branch module. (2) By distinguishing the positive and negative sample pairs in each batch, the TRANSLATOR modules in the two branches are forced to learn the information of their respective modalities, so as to realize the goal of learning the information of two modalities at the same time. In the proposed method, the last hidden state of the TRANSLATOR in each branch are used to compute the contrastive loss. The last hidden state is the hidden state of the last time step of the TRANSLATOR's last layer decoder, which has the information of all time steps simultaneously. Compared with the dimension reduction of the two modalities features by linear transformation, the direct use of last hidden state can avoid the information loss caused by dimension reduction. The way to calculate the contrastive loss is similar to CLIP\cite{radford2021clip}, which can be expressed as follows:
\[L_{a\rightarrow c} =\frac{1}{N}  {\textstyle \sum_{i=1}^{N}} \log{\frac{\exp \left ( h_{i}^{a}  \cdot h_{i}^{c}  \right ) }{ {\textstyle \sum_{j=1}^{N}} \exp \left ( h_{j}^{a}  \cdot h_{j}^{c} \right ) } }\tag{5}\]
\[L_{c\rightarrow a} =\frac{1}{N}  {\textstyle \sum_{i=1}^{N}} \log{\frac{\exp \left ( h_{i}^{a}  \cdot h_{i}^{c}  \right ) }{ {\textstyle \sum_{j=1}^{N}} \exp \left ( h_{j}^{c}  \cdot h_{j}^{a} \right ) } }\tag{6}\]
\[L_{cl}  = \frac{1}{2} \left ( L_{a\rightarrow c} \ +\ L_{c\rightarrow a}  \right ) \tag{7}\]
Where \(N\) denotes the batch size, \(\left ( h^{a} ,\ h^{c} \right )\) denotes the TRANSLATOR's last hidden state in the audio branch and text branch, respectively. 
The total loss of the model can be expressed as follows:
\[L = L_{ce} \ + \ L_{cl} \tag{8}\]

\section{Results}

This section will introduce the experimental procedure, training strategy and evaluation metrics in detail

\subsection{Dataset}

We verify the feasibility of the proposed method in Clotho V2\cite{drossos2020clotho} and AudioCaps\cite{kim2019audiocaps}, respectively.

\begin{table*}[]
\large
\caption{Results of ablation experiments on the Clotho\cite{drossos2020clotho} evaluation split. The effectiveness of the method is verified by adding modules one by one on the basis of baseline. Where Encoders means that three audio encoders (CNN14\cite{kong2020panns}, HTSAT\cite{chen2022hts}, ConvNeXt\cite{liu2022convnxet}) are used at the same time, and (Sim) TRANSLATOR means that twin TRANSLATOR structure is used. All encoder weights are frozen during training.}\label{tab:tablenotes}
\renewcommand\arraystretch{1.3}

\setlength{\tabcolsep}{3mm}{\resizebox{\textwidth}{!}{
\begin{tabular}{c|c|ccccc}
\hline
\textbf{Adding modules}           & \textbf{Module composition}                  & \textbf{METEOR} & \textbf{CIDER} & \textbf{SPICE} & \textbf{SPIDEr} & \textbf{SPIDEr\_FL} \\ \hline
\multirow{3}{*}{baseline}         & CNN14 + BART                                & 0.169  & 0.405 & 0.115 & 0.260  & 0.257     \\
                                  & HTSAT + BART                                & 0.178  & 0.426 & 0.126 & 0.276  & 0.274     \\
                                  & ConvNeXt + BART                             & 0.172  & 0.398 & 0.122 & 0.260  & 0.256     \\ \hline
\multirow{2}{*}{FUSER}            & Encoders   + FUSER + BART                   & 0.185  & 0.463 & 0.131 & 0.297  & 0.293     \\
                                  & HTSAT +   FUSER + BART                      & 0.184  & 0.454 & 0.133 & 0.293  & 0.292     \\ \hline
\multirow{2}{*}{TRANSLATOR}       & Encoders   + FUSER + TRANSLATOR + BART      & 0.186  & 0.470 & 0.135 & 0.302  & 0.297     \\
                                  & HTSAT +   TRANSLATOR + BART                 & 0.182  & 0.450 & 0.133 & 0.290  & 0.288     \\ \hline
\multirow{2}{*}{Twin   structure} & HTSAT +   (Sim) TRANSLATOR + BART            & 0.182  & 0.453 & 0.132 & 0.293  & 0.286     \\
                                  & Encoders   + FUSER + (Sim) TRANSLATOR + BART & 0.188  & 0.483 & 0.135 & 0.309  & 0.304     \\ \hline 
\end{tabular}}}
\end{table*}

\subsubsection{Clotho V2}

Clotho V2\cite{drossos2020clotho} is split into development, Evaluation, validation, and test splits. In addition to the test split, the other three splits are all composed of audio samples with a duration of 15 to 30 seconds, and each audio sample contains five captions of length 5-10. As the official dataset for the DCASE challenge, there are a total of 6974 audio samples with 34870 captions.

\subsubsection{AudioCaps}

The AudioCaps\cite{kim2019audiocaps} data consists of 50K audio samples picked by Audioset\cite{gemmeke2017audioset} and split into training, validation, and test splits. The length of each audio sample is 10s. Each audio sample corresponds to one caption in training splits, and each audio sample corresponds to five captions in validation and test splits.

\subsection{Evaluation metrics}

The evaluation metrics for AAC tasks can be divided into two categories, which are based on n-gram overlap and based on semantic similarity\cite{mei2022automated}. BLEU-n\cite{papineni2002bleu} evaluates the grammatical match between the predicted value and the ground truth value by calculating n-grams, but there are problems where short captions are more likely to obtain high scores. ROUGE-L\cite{chin2004rouge} is similar to BLEU but focuses more on recall. METEOR\cite{banerjee2005meteor} is more focused on calculating the F1 score and recall for individual words. CIDEr\cite{vedantam2015cider} weights n-grams by calculating TF-IDF scores. SPICE\cite{anderson2016spice} is used to evaluate how well the predicted values match the ground truth semantically. SPIDEr\cite{liu2017spider} is the average of SPICE and CIDEr. SPIDEr-FL is a newly proposed metric, which is obtained by calculating the average of SPIDEr and FENSE\cite{zhou2022fense}.

Since the start of the DCASE 2023 challenge, METEOR, CIDEr, SPICE, SPIDEr, and SPIDEr-FL have been used as metrics for AAC tasks. We use the same metrics as DCASE to evaluate the proposed method, and other metrics will also be listed for reference.

\subsection{Experimental environment and parameter Settings}

We use hyperparameter settings that have been extensively validated in previous studies. AdamW is used as the optimizer in the training process with batch size 32. The initial learning rate was \(3\times 10^{-5} \), and the cosine annealing algorithm was used to update the learning rate. In a cycle of 4 epochs, the benchmark learning rate first decreased by 10 times and then recovered in a cycle. Each epoch is evaluated three times, each time comparing the CIDEr\cite{vedantam2015cider} to determine whether the weight is saved. In addition, label smooth is applied to reduce the problem of model overconfidence. Use teacher forced methods to shorten the training time. Each audio encoder uses SpecAugment\cite{park2019specaugment} for data augmentation.

We train for 20 epochs on a single NVIDIA V100 (32GB) GPU, halve the baseline learning rate every 4 epochs. The strategy of early stopping is used to end the training when the metrics are not updated for 4 consecutive epochs.

In the inference phase, the beam search method with num\_beams = 4 is used to generate captions, and data augmentation is disabled.

\subsection{Comparison with other methods}

We compare the proposed method with the current best performing method, as shown in Table 1. Even though BLEU\cite{papineni2002bleu} and ROUGE\cite{chin2004rouge} have been removed from the final results since the DCASE 2023 challenge, for the sake of fairness, we still list them for reference.

Our method outperforms the current state of the art methods on the AudioCaps\cite{kim2019audiocaps} dataset and achieves competitive results with the SOTA method on the Clotho\cite{drossos2020clotho} dataset. Notably, Mei et al.\cite{mei2023wavcaps} 's method underperforms our method on the AudioCaps dataset even pre-trained on the WavCaps\cite{mei2023wavcaps} dataset (which is almost 60 times larger than Clotho), while it achieves only a slight advantage \( \left ( \le 0.005 \right ) \) on the Clotho dataset.

We also specifically compare with the method proposed by Wu et al.\cite{wu2023SOTA}, as shown in Table 3, which achieves the state of the art performance on the Clotho dataset. Wu et al.\cite{wu2023SOTA} 's method uses the AudioCaps dataset for pre-training and ChatGPT to assist in integrating captions in the dataset, which essentially uses additional data for training. The results show that our method, both without and with text embeddings, performs better on the Clotho dataset than the method proposed by Mei et re\cite{wu2023SOTA}, despite not training with additional data. And the number of parameters is only about 35\% of Wu et al.\cite{wu2023SOTA}

It is important to point out that our approach is \textbf{not} fine-tune with reinforcement learning.\cite{mei2021rl} Because it will improve the metrics by repeatedly generating keywords which will reduce the fluency of the generated captions.

\subsection{Ablation Study}

We conduct an ablation study of the proposed method on the Clotho\cite{drossos2020clotho} dataset by adding a single module to the baseline model benchmark. The specific results are shown in Table 2. The architecture of a single audio encoder and using BART\cite{lewis2019bart} as the text decoder is the baseline model. During training, the weights of all encoders are frozen.

Lines 1-3 show that HTSAT\cite{chen2022hts} performs best when BART is used as the text decoder. We tried using other audio encoders (BEATs\cite{chen2022beats}, etc.) but didn't get better results. Lines 4-5 show that adding the FUSER module proposed in this paper can achieve significant results. This means that the FUSER module can effectively combine different audio features. The results in line 5 show that efficient results are still achieved when using a single audio encoder as input to the FUSER module, but the performance is not as good as using multiple audio encoders. This indicates that the FUSER module can also propose higher dimensional semantic information from a single audio features.

The results in lines 6-9 show that both the TRANSLATOR module and the twin structure proposed in this paper can achieve effective results, but the boost is less than that of the FUSER module. It is worth noting that under the proposed architecture, the TRANSLATOR module must be used when using the twin structure. We tried using InfoNCE\cite{he2020MoCo} loss as the contrastive loss, but it didn't perform as well as CLIP loss.

\begin{table}[]
\Huge
\caption{Detailed comparison with SOTA\cite{wu2023SOTA}. The baseline representation only uses audio embeddings. text embs means adding text embeddings to the baseline model. Rerank represents the reordering of captions using additional methods.}\label{tab:tablenotes}
\renewcommand\arraystretch{1.3}
\setlength{\tabcolsep}{1mm}{\resizebox{\columnwidth}{!}{
\begin{tabular}{c|c|ccccc}
\hline
\textbf{Modules}       & \textbf{}                  & \textbf{METEOR} & \textbf{CIDER} & \textbf{SPICE} & \textbf{SPIDEr} & \textbf{SPIDEr\_FL} \\ \hline
\multirow{2}{*}{baseline}                                                                  & Ours & 0.186 & \textbf{0.470} & \textbf{0.135} & \textbf{0.302} & \textbf{0.297} \\
                                                                                                   & \huge{SOTA} & 0.186          & 0.458          & 0.134          & 0.296          & 0.294          \\ \hline
\multirow{2}{*}{\begin{tabular}[c]{@{}c@{}} + \\text embs\end{tabular}} & Ours & \textbf{0.188} & \textbf{0.483} & \textbf{0.135} & \textbf{0.309} & \textbf{0.304} \\
                                                                                                   & \huge{SOTA} & 0.187          & 0.474          & 0.134          & 0.304          & 0.303          \\ \hline
\multirow{2}{*}{\begin{tabular}[c]{@{}c@{}} + \\Rerank\end{tabular}} & Ours & \textbf{-} & \textbf{-} & \textbf{-} & \textbf{-} & \textbf{-}  \\  & \huge{SOTA} & 0.193          & 0.506          & 0.146          & 0.326          & 0.326          \\ \hline
\end{tabular}}}
\end{table}

\section{Conclusion and future work}

In this work, in order to overcome the disparity between the two modalities, three perspectives are proposed to jointly assist the model to understand the text information. Firstly, FUSER is proposed to extract common semantic information from different audio features by feature fusion. Then, by proposing TRANSLATOR, audio features are aligned to text features from the tensor level. Finally, through the twin structure, the model in the audio branch can retain its ability to extract audio features, and learn text information from text features at the same time, so as to make use of text information as efficiently as possible. The FUSER module and TRANSLATOR module proposed in this study can be used as plug-and-play semantic feature extractors and applied to almost models in AAC research.

In future work, self-supervised methods may be combined to try not to use additional text encoders, which will fundamentally solve the problem of resource waste due to the introduction of additional encoders.


\end{document}